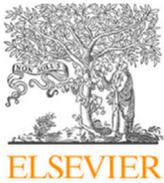

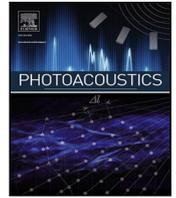

Research article

# Photoacoustic image synthesis with generative adversarial networks

Melanie Schellenberg [a,b,c,*], Janek Gröhl [a,d], Kris K. Dreher [a,e], Jan-Hinrich Nölke [a,b], Niklas Holzwarth [a,b], Minu D. Tizabi [a], Alexander Seitel [a], Lena Maier-Hein [a,b,c,f,g,*]

[a] Intelligent Medical Systems (IMSY), German Cancer Research Center (DKFZ), Heidelberg, Germany
[b] Faculty of Mathematics and Computer Science, Heidelberg University, Heidelberg, Germany
[c] HIDSS4Health - Helmholtz Information and Data Science School for Health, Heidelberg, Germany
[d] Cancer Research UK Cambridge Institute, University of Cambridge, Robinson Way, Cambridge, CB2 0RE, UK
[e] Faculty of Physics and Astronomy, Heidelberg University, Heidelberg, Germany
[f] Medical Faculty, Heidelberg University, Heidelberg, Germany
[g] HIP Applied Computer Vision Lab, Medical Image Computing, German Cancer Research Center (DKFZ), Heidelberg, Germany

## ARTICLE INFO

## ABSTRACT

Photoacoustic tomography (PAT) has the potential to recover morphological and functional tissue properties with high spatial resolution. However, previous attempts to solve the optical inverse problem with supervised machine learning were hampered by the absence of labeled reference data. While this bottleneck has been tackled by simulating training data, the domain gap between real and simulated images remains an unsolved challenge. We propose a novel approach to PAT image synthesis that involves subdividing the challenge of generating plausible simulations into two disjoint problems: (1) Probabilistic generation of realistic tissue morphology, and (2) pixel-wise assignment of corresponding optical and acoustic properties. The former is achieved with Generative Adversarial Networks (GANs) trained on semantically annotated medical imaging data. According to a validation study on a downstream task our approach yields more realistic synthetic images than the traditional model-based approach and could therefore become a fundamental step for deep learning-based quantitative PAT (qPAT).

## 1. Introduction

Multispectral photoacoustic tomography (PAT) is an emerging medical imaging modality that provides morphological and functional tissue information with high contrast and spatial resolution in tissue depths up to several centimeters [1,2]. Despite recent successes, numerous PAT applications are not yet feasible in clinical settings [3]. One of the primary obstacles related to clinical translation of PAT in these scenarios is that the absolute concentration of different absorbers, referred to as chromophores, cannot be quantified reliably. Even though the initial pressure distribution reconstructed from measured PAT raw data is proportional to the optical absorption coefficient of chromophores, it is also proportional to the light fluence which, in turn, is highly dependent on the distribution of chromophores and scatterers within the tissue. Consequently, quantification of chromophore concentrations from the underlying initial pressure distribution is an ill-posed inverse problem (*optical inverse problem*). Current model-based quantification approaches do not yield accurate results in clinical settings, mainly due to assumptions that do not hold in the clinical context [4,5]. Machine learning-based solutions have recently been proposed as an alternative approach to model-based quantitative PAT (qPAT) [6–8]. Here, the absence of a reliable reference method for generating annotated training data has been addressed by simulating training data with ground truth-underlying tissue properties. While these works obtained very encouraging results *in silico*, the data-driven approaches to qPAT tend to fail in realistic scenarios, among others due to the domain gap between real and simulated data [6,9].

Various solutions have been developed to address this challenge in the field of general machine learning [10]. Domain adaptation is a particular case of transfer learning that utilizes labeled data in one or multiple source domains (here: simulated data) to execute new tasks on data of a target domain [11] (here: real device data). The approach has successfully been applied in the field of biomedical imaging data, such as multispectral [12], ultrasound (US) [13], and PAT [14,15] imaging data, for example.

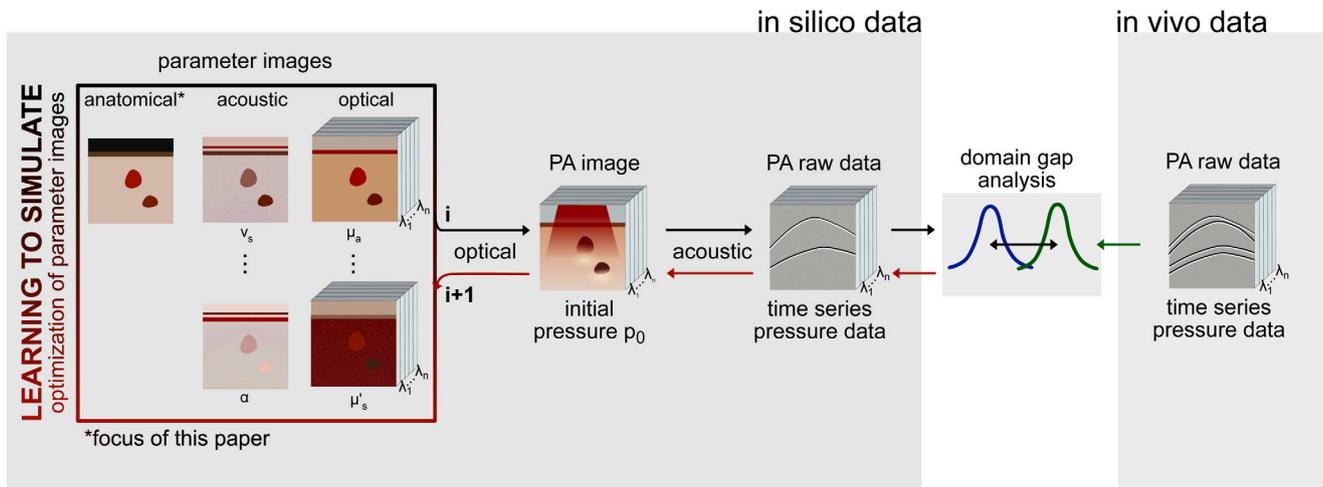

**Fig. 1.** Our data-driven approach to quantitative photoacoustic tomography (SPICE-qPAT[1]): We consider image formation (black arrows) and image decoding (red arrows) in one joint framework. The core of our concept is the explicit disentanglement of the different factors that contribute to image formation. These include anatomical, acoustic (e.g., the speed of sound $v_s$ and the attenuation coefficient $\alpha$) and optical tissue parameters (e.g., the absorption coefficient $\mu_a$ and reduced scattering coefficient $\mu_s'$ at wavelength $\lambda_i$), all of which are represented by *parameter images*. The optical inverse problem is phrased as the recovery of these parameter images from the initial pressure images. The acoustic inverse problem relates to the reconstruction of the initial pressure images from the recorded raw time series pressure data. Our concept involves a machine learning-driven approach for the generation of the parameter images to close the domain gap between simulated and real measurements. Within this broader context, the current paper focuses on the generation of plausible tissue geometries, as detailed in Fig. 2.

While domain adaptation is a general method used to address distribution shifts between training and test data, other approaches have been specifically designed for simulated data. One way to enhance the realism of simulated data is to replace existing, often non-differentiable, model-based simulations by learnable and differentiable components [16,17]. The "realistic" data generated by these methods are widely adopted as data augmentation in various tasks, especially precious in the biomedical field [18–25]. Another approach to increase the realism of simulations is the extension of model-based simulations with learnable components [26–28]. As an example, simulation parameters may be optimized by reinforcement learning-based methods, while the simulations as such remain model-based [28].

Overall, we observe an increasing scientific interest in the topic of learning from simulations, but in the specific context of qPAT we are not aware of any work that uses machine learning concepts to substitute the forward image generation process or to tune the simulation parameters more realistically.

We address this gap in the literature with a novel method, referred to as SPICE-qPAT.[1] SPICE-qPAT systematically combines established knowledge from biomedicine and physics with modern machine learning concepts to enable qPAT. The concept is based on two core components: A physics-based simulation framework for virtual spectral imaging and a neural network-based decoding framework that leverages both simulated data and real data to learn the recovery of relevant tissue properties from measurement data. Here, image synthesis and decoding are considered in one joint framework that explicitly disentangles all the different optical, acoustic, and geometric tissue properties that contribute to image formation (cf. Fig. 1).

Within this broader research context, the contribution of this paper focuses on the realistic simulation of PAT data. Concretely, we present a simulation concept that leverages real medical imaging data to optimize the realism of tissue geometries that serve as a basis for PAT image synthesis (cf. *anatomical* parameter images in Fig. 1). The following sections present our approach to synthetically generate tissue geometries as well as our initial feasibility study to validate this approach.

## 2. Material and methods

The ensuing sections present the proposed framework for "learning to simulate" (cf. Section 2.1), the specific approach to data-driven learning of plausible tissue geometries (cf. Section 2.2), as well as the initial feasibility study conducted to validate the latter (cf. Section 2.3).

### 2.1. "Learning to simulate" framework

In our approach, the problem of qPAT is formulated as a decoding task in which neural networks are applied to convert measurements pixel-wise (in 3D also referred to as voxel-wise) to underlying parameter images. In this context, image synthesis and decomposition are considered together as one joint problem. While previous approaches to deep learning-based qPAT have focused on solving the quantification problem directly, our approach relies on the explicit disentanglement and an analysis of the direct dependencies of all of the parameters that are relevant for the image formation (cf. Fig. 1). These parameters are represented by *parameter images* and comprise three classes: anatomical, optical, and acoustic parameters. The anatomical parameters describe the anatomy of different tissue types and therefore the spatially-resolved concentration of chromophores. Optical parameters, such as the absorption ($\mu_a$) and scattering ($\mu_s$) coefficients and acoustic parameters, such as the speed of sound ($v_s$), specify the molecular information relevant for the optical and acoustic image formation process, respectively.

A core component of SPICE-qPAT is the simulation framework. While previous simulation approaches have been purely model-based [29], a key feature of our concept is the fact that we leverage real data to learn parameter images in a data-driven manner. More specifically, we focus on the generation of plausible anatomical parameters to quantify the optical properties from the PA image (*optical inverse problem*), as detailed in the following sections.

### 2.2. Learning tissue geometries

Recent success stories in the field of machine learning [30–32] have shown the high potential of Generative Adversarial Networks (GANs) [33] and their numerous variants in synthesizing realistic content. In the present work, we leverage GANs for synthetic PAT image generation in the following 6-step process (cf. Fig. 2).

---

[1] Named after the corresponding European Research Council (ERC) grant "NEURAL SPICING" (short: SPICE).

## i) GAN-based generation of parameter images

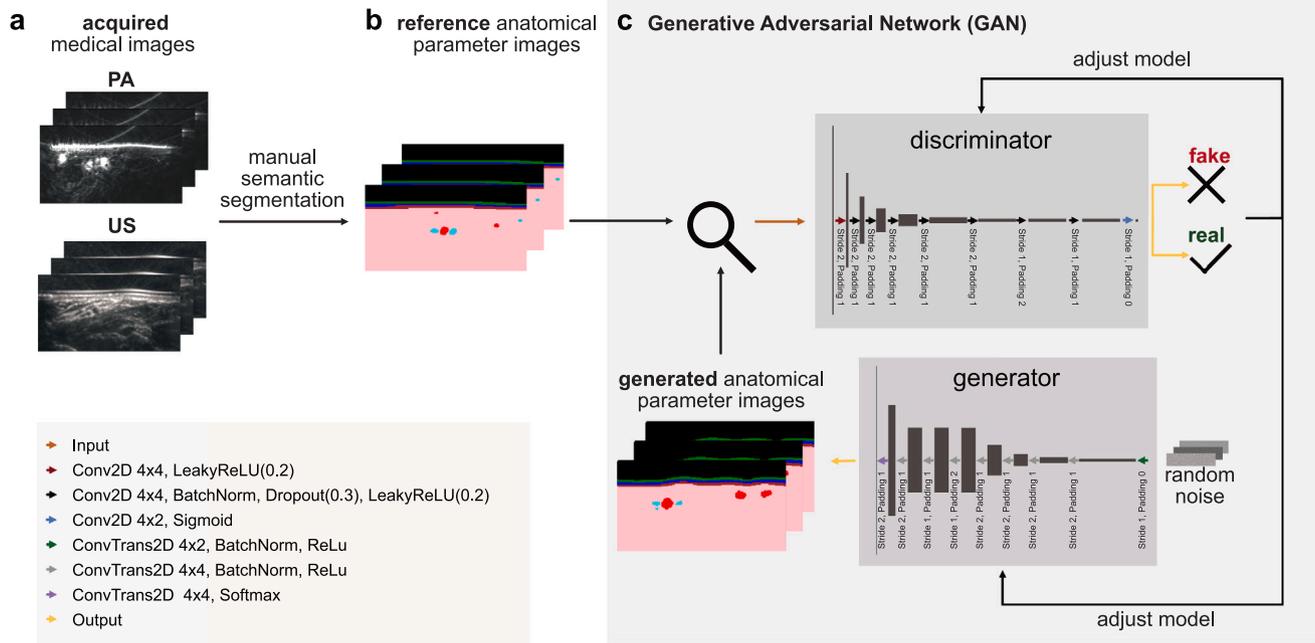

## ii) From parameter images to synthetic PAT images

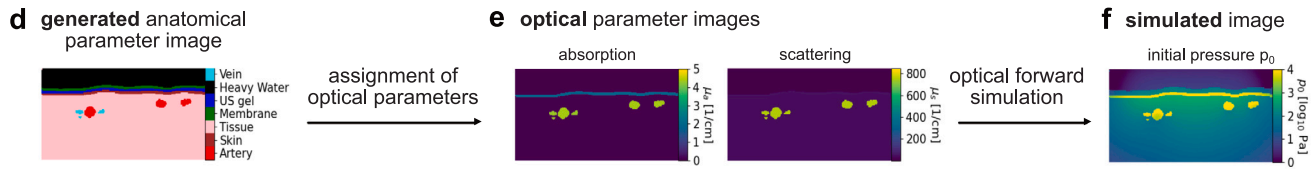

**Fig. 2.** Concept for data-driven generation of synthetic photoacoustic tomography (PAT) training data. **(i)** The generative adversarial network (GAN)-based generation of *anatomical parameter images*. **(a)** Medical images related to the tissue geometry (here PAT co-registered to ultrasound (US) data) are semantically segmented. These **(b)** reference *anatomical parameter images* are used to train a **(c)** GAN for the generation of anatomical parameter images. While the generator network learns to generate realistic parameter images representing tissue geometry, the discriminator network learns to distinguish real from fake ones. The optimization of the competing networks leads to generated realistic segmentation masks with a data distribution identical to the data distribution of real segmentation masks. **(ii)** The simulation of PAT images from the parameter images. Leveraging **(d)** a generated anatomical parameter image and **(e)** the corresponding optical (and depending on the application also acoustic) parameter image, **(f)** a new training image can be simulated.

(a) **Image acquisition:** Acquisition/Gathering of tomographic 2D or 3D images of the target anatomy with any modality, such as computed tomography (CT), magnetic resonance imaging (MRI), US, and PAT.

(b) **Image annotation:** Generation of semantic segmentations reflecting the format of the desired anatomical parameter images (cf. Fig. 2 for an example). Specifically, the input images are classified pixel-wise in a way that different classes present structures with systematically different acoustic and/or optical properties. In the case of forearm images, for example, the tissue classes veins, arteries, skin, and background tissue are labeled.

(c) **Training of Generative Adversarial Network (GAN):** A GAN is trained to generate anatomical parameter images, resembling those in the training data set. Generally, a GAN consists of two networks, a generator and a discriminator network [33]. The generator network enables the generation of fake data. In particular, it learns to map a low-dimensional latent space, such as Gaussian distributed random noise, to a high-dimensional output, such as high-resolution synthetic images, with a data distribution identical to the real training data distribution. In contrast, the discriminator network is a classifier that learns to distinguish the real data from fake data. As the networks are competitors that continuously trigger mutual improvement, the optimization of both networks leads to generated realistic synthetic data. In our current framework, we apply the deep convolutional GAN [34] as it is a particularly preferred architecture for image synthesis [35].

(d) **Image generation:** The trained GAN is used to generate any number of plausible anatomical parameter images (semantic segmentation maps).

(e) **Generation of optical parameter images:** Based on the geometrical information, the remaining (here optical) parameter images are generated (typically also in a probabilistic manner) as described in Section 2.3.4.

(f) **Generation of PAT images:** The optical parameter images were used as the input of our simulation pipeline [36].

The proposed workflow allows leveraging geometric information accessible from other modalities and directly addresses the fact that tissue geometries are relatively easy to label, while there is no reliable reference method for assigning optical and acoustic properties. The following section presents the first feasibility study we conducted to investigate the potential of the approach for deep learning-based solving of the optical inverse problem. Implementation details of our first prototype can be found in Section 2.3.2.

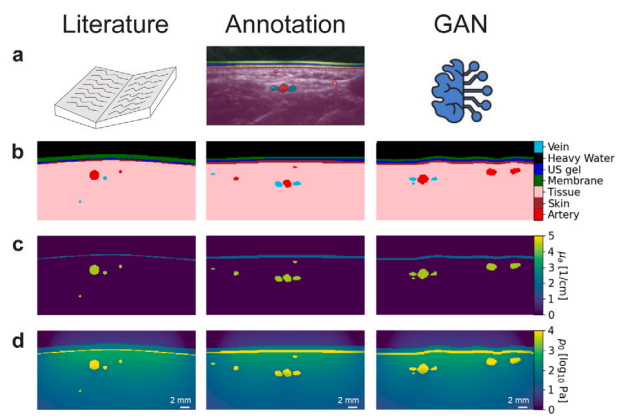

**Fig. 3.** Examples of the *(left)* literature-based, *(center)* annotation-based, and *(right)* Generative Adversarial Network (GAN)-based forearm parameter images. Leveraging **(a)** the respective anatomical information source, **(b)** the anatomical parameter images are generated and used to assign **(c)** the corresponding optical tissue parameters of a human according to literature values (e.g., the absorption coefficient $\mu_a$ shown) which enables the simulation of **(d)** the initial pressure distribution $p_0$ ($\mu_a$ and $p_0$ here shown at 800 nm).

**Table 1**
Hyperparameters used for Generative Adversarial Network (GAN) training.

| Hyperparameter | Value | | |
|---|---|---|---|
| **General** | | | |
| Max epochs | 700 | | |
| Batch size | 3 | | |
| Label smoothing $P_{smooth}$ | $\mathcal{U}(-0.3, 0.0)$ | | |
| Label flipping $P_{flip}$ | Decrease with epochs | | |
| intercept($P_{flip}$) | 0.2 | | |
| slope ($P_{flip}$) | $-2.9 \cdot 10^{-4}$ | | |
| **Augmentation of real and fake images [40]** | | | |
| $P_{trans}$ per affine transformation | 0.6 | | |
| rotation | $\mathcal{U}(-45°, 45°)$ | | |
| x-translation | $\mathcal{U}(-5\ pix, 5\ pix)$ | | |
| y-translation | $\mathcal{U}(-5\ pix, 5\ pix)$ | | |
| **Generator** | | | |
| Learning rate | $2 \cdot 10^{-4}$ | | |
| | Forearm | Calf | Neck |
| Initial channel size | 100 | 106 | 100 |
| **Discriminator** | | | |
| Learning rate | $2 \cdot 10^{-4}$ | | |
| Initial channel size | 56 | | |

$\mathcal{U}(A,B)$ = uniform distribution between A and B
$P$ = probability

## 2.3. Experiments

The purpose of the experiments was to assess the benefit of our data-driven approach to the generation of tissue geometries of three different body sites, namely forearm, calf, and neck. We compared our method (*GAN-based*) to an approach directly utilizing annotated semantic segmentation masks of real PAT measurements (*annotation-based*). In particular for forearm data, we additionally compared our method to an approach leveraging literature knowledge for model-based generation of tissue geometries (*literature-based*). Note that in contrast to forearms, major vessels in calves and necks are generally located deeper within the tissue [37]. Given the imaging depth of our PAT device ($\sim 2$ cm), only superficial vessels can be imaged for these body sites. To our knowledge, there is a lack of literature for the anatomical properties of these vessels, which hindered us in setting up respective model-based approaches.

The following sections present the different approaches for generating tissue geometries – namely the reference annotation-based approach (cf. Section 2.3.1), our GAN-based approach (cf. Section 2.3.2), and the baseline literature-based approach for forearm data (cf. Section 2.3.3) – as well as the simulation pipeline (cf. Section 2.3.4) and our strategy for comparative validation (cf. Section 2.3.5).

### 2.3.1. Annotation-based generation of tissue geometries

As a reference to realistic PAT segmentation masks, we used semantic segmentation masks of 96 pairs of US and PAT images of 16 healthy human volunteers per body site (forearm, calf, neck). These image pairs were acquired using the MSOT Acuity Echo device (iThera Medical, Munich, Germany) with the consent of all volunteers and in compliance with relevant regulations (German Clinical Trials Register reference number DRKS00023205). Freehand scans of roughly 30 s at three positions at the right and left forearm of every volunteer were acquired following an acquisition protocol [38]. While the US images were reconstructed using a backprojection algorithm by the MSOT device itself, the PAT images were reconstructed using a Delay-And-Sum (DAS) algorithm implemented within the Medical Imaging Interaction Toolkit (MITK) [39]. The PAT images were post-processed in four steps. First, the multispectral PAT images were corrected for laser pulse energy variations of the MSOT device. Each pixel value of the PAT image was divided by the respective laser pulse energy. Second, to account for the different fields of view of PAT and US resulting from the different reconstruction algorithms used, the PAT images were cropped in depth direction, such that a co-registration with the US images was possible. Third, the image pairs were re-sampled to have an isotropic spacing of 0.16 mm and, fourth, the pairs were divided into four sub-scans of approximately eight seconds each. Every sub-scan was averaged pixel-wise and the resulting image pair with the sharpest edges in the US image according to the averaged calculated image gradient was selected. Following a detailed annotation protocol, which can be found in the supplemental material of [38], these selected image pairs were semantically segmented by domain experts into the following classes: (1) artery, (2) skin, (3) muscle background tissue, (4) US gel, (5) transducer membrane, (6) heavy water of the transducer head, and (7) vein. These reference semantic segmentation masks were intended to include realistic anatomical parameters. Fig. 3 (*annotation* column, b–d) shows a randomly chosen example of the annotation-based forearm data set.

### 2.3.2. GAN-based generation of tissue geometries

While the annotation-based approach potentially yields the most accurate tissue geometries, it does not scale well due to the need for manual annotations. To compensate for the data sparsity, we propose a GAN-based approach that generates further plausible geometries by leveraging existing morphological imaging data (not necessarily acquired with PAT). Based on the presented annotation-based data set (cf. Section 2.3.1), the proposed concept for data-driven tissue generation was implemented as follows: First, the training data comprising 78 annotated images per body site (not included in any test data) was augmented by horizontally flipping a copy of each image and adding it to the training set, which was thereby doubled in size. Then, a deep convolutional GAN architecture shown in the top part of Fig. 2 was trained on that body site specific data. In other words, one GAN was trained on forearm data only, one on calf data only, and one on neck data only. To deal with the limited amount of data, we followed [40] and augmented the training data with rotation and translation before the discriminator was applied. The hyperparameters (cf. Table 1) were determined by applying a grid search on the training data. With this generative model trained, 500 diverse anatomical parameter images were generated per body site. Fig. 3 (*GAN* column, b–d) shows a randomly chosen example of this GAN-based forearm data set.

**Table 2**
Different configurations of three data sets per body site (forearm (**a–e**), calf (**a–c**), and neck (**a–c**)) with respective training, validation, and test split used for the U-Net-based quantitative downstream task. The target test data set is highlighted in bold green.

| Data set | Training 70% | Validation 10% | Test 20% |
|---|---|---|---|
| **a** annotation-based (anno) | 66 | 12 | **18** |
| **b** Generative Adversarial Network-based (GAN) | 350 | 50 | 100 |
| **c** combination of **a** and **b** (GAN-anno) (ratio: $\frac{a}{b} = \frac{19}{81}$) | 350 | 50 | 100 |
| **d** *forearm only:* literature-based (lit) | 350 | 50 | 100 |
| **e** *forearm only:* combination of **a**, **b**, and **d** (lit-GAN-anno) | 766 | 112 | 218 |

### 2.3.3. Literature-based generation of forearm tissue geometries

The literature-based tissue geometries of anatomical tissue parameters of different tissue classes of a human forearm including epidermis, muscle tissue, arteries, and veins, are based on an internally developed forearm tissue model in previous work [29,41]. We leveraged this model to generate 500 literature-based tissue geometries. Further details of the literature model are specified in Section D in the supplemental material. Fig. 3 (*literature* column, b–d) shows a randomly chosen semantic segmentation mask of the literature-based forearm data set.

### 2.3.4. Photoacoustic image synthesis

Based on the anatomical parameter images, the optical parameter images were generated. To this end, the internal tissue library of the open-source "Simulation and Image Processing for Photonics and Acoustics" (SIMPA)[2] [36] toolkit, which contains the optical parameters for our annotation classes, was used. Therefore, similar to [42], the optical parameters were assigned per tissue class. Since the optical parameters for vessels and tissue depend on the blood oxygenation ($sO_2$) and the blood volume fraction, they were defined indirectly by sampling from the distributions provided in Table D.1 b in the supplemental material. Multispectral images were simulated in 16 wavelengths ranging from 700–850 nm in steps of 10 nm leveraging the Monte Carlo model implementation "Monte Carlo eXtreme" [43] included in the SIMPA toolkit using the same simulation properties for all data sets (cf. bottom part in Fig. 2). The different tissue geometries were placed in the simulation volume such that the highest point of the ultrasound gel layer had a constant distance of 43.2 mm to the upper border of the simulation volume to account for the in-plane distance to the illumination source of the MSOT Acuity Echo probe. To obtain the requested input format of the SIMPA toolkit, which performs simulations in 3D, the 2D annotation- and GAN-based segmentation masks were stacked along the y-axis and extrapolated along the x- and z-axes to match the dimensions of the simulation volume. After simulation of all data sets the final 2D simulated images were chosen as the center x-z slice of the simulation volume. Further details for the simulation process can be found in Table D.1 (a and b) in the supplemental material.

### 2.3.5. Comparative performance assessment

To investigate the benefit of our "learning-to-simulate" approach, we assessed the effect of the method for geometry generation on a downstream task, namely the reconstruction of the spatially-resolved optical absorption $\mu_a$ from PAT images. Both the PAT and the optical absorption parameter images were scaled logarithmically and Gaussian noise ($\mu = 0$, $\sigma = 0.5$) was added to the PA images. Our strategy involved training qPAT models, all of the same U-Net architecture (cf. Fig. B.1 in the supplemental material), on data sets corresponding to the strategies presented in Sections 2.3.1–2.3.3 (annotation-based, GAN-based, literature-based), and combinations thereof. More specifically, we used three different configurations per body site and two additional ones for the forearm data shown in Table 2 and subdivided the data into training (70%), validation (10%), and test data (20%).

Note that the same annotations were used as a basis for the annotation-based and GAN-based approaches. The GAN-based approach can thus be seen as a form of data augmentation.

Previous works [44] and a recent review on deep learning for biomedical PAT [6] have demonstrated that the U-Net architecture shows particular promise in tackling the optical inverse problem. The U-Net architecture applied is shown in Figure B.1 in the supplemental material. All 16 wavelengths were leveraged simultaneously, meaning the input and output dimensions of the U-Net were set to $256 \times 128 \times 16$ where the first two dimensions define the image size and the last dimension represents the 16 wavelengths. The hyperparameters were determined by a grid search analysis on the corresponding validation data set (cf. Tab. B.1 in the supplemental material). Test results were then determined on both the respective held-out test set as well as on the realistic annotation-based test set highlighted in green in Table 2.

A quantitative analysis of the estimated absorption coefficients $\hat{\mu}_a$ on the respective test data set $x$ was performed per test image and separately for different tissue classes (cf. Section 2.3.1). The relative error, $RE_{x,i_c,\lambda}$, the absolute error $AE_{x,i_c,\lambda}$, and the structural similarity index ($SSIM$) [45] were used. The relative and absolute errors are defined as:

$$RE_{x,i_c,\lambda} = \frac{|\hat{\mu}_{a,i_c,\lambda} - \mu_{a,i_c,\lambda}|}{\mu_{a,i_c,\lambda}}, \tag{1}$$

$$AE_{x,i_c,\lambda} = |\hat{\mu}_{a,i_c,\lambda} - \mu_{a,i_c,\lambda}|, \tag{2}$$

where $x$ is the test data index and $i_c$ is the pixel index in class $c\{1,...,7\}$ denoting artery, skin, muscle tissue, transducer membrane, US gel, transducer head mainly consisting of heavy water, and vein, respectively. $\lambda$ is the evaluated wavelength and $\hat{\mu}_{a,i_c,\lambda}$ and $\mu_{a,i_c,\lambda}$ are the estimated and ground truth (GT) absorption coefficient at pixel $i_c$ and wavelength $\lambda$, respectively.

For comparison of the different techniques, we applied the *challengeR* method [46]. It is especially suited for analyzing and visualizing challenge results across different tasks in the field of biomedical image analysis. The challenge in our context is the performance of the downstream task. The competing algorithms were defined as the three/five algorithms corresponding to the training data sets presented in Table 2 (all processed with the same neural network model and tested on the target test data set). We then used the tool to compute performance images separately for the three metrics and different tissue classes. As *tasks*, we defined the quantification of optical absorption for the different wavelengths (n = 16 in total) and the *test cases* corresponded to the 18 test data sets. In the $SSIM$ scenario, the $SSIM$ results were chosen directly as the *values*. For the other scenarios, *values* were defined by averaging the pixel-wise metrics $AE_{x,i_c,\lambda}$ and $RE_{x,i_c,\lambda}$ over the tissue classes corresponding to $\overline{AE}_{x,c,\lambda}$ and $\overline{RE}_{x,c,\lambda}$. Additionally, the per-class metric values were averaged to give an "overall" measure denoted with $c = 0$. Target classes were defined as the classes artery, skin, and vein (c = 1, 2, 7). In particular, the tool was used in the rank-then-aggregate mode to investigate the consensus ranking stability (default settings with mean aggregation).

## 3. Results

The performance of the U-Net-based models trained on different configurations of the three data sets (cf. Table 2) is presented in the following sections for the three body sites: forearm (cf. Section 3.1), calf and neck (cf. Section 3.2).

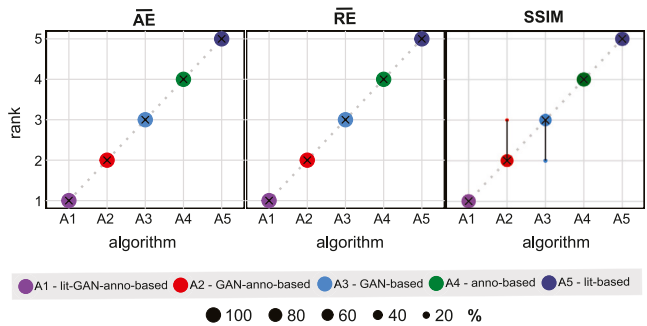

**Fig. 4.** Comparative performance assessment of the forearm models corresponding to different training sets (cf. Table 2) and tested on identical annotation-based test data. Uncertainty-aware rankings were computed for the mean absolute error ($\overline{AE}_{c,\epsilon 0}$), mean relative error ($\overline{RE}_{c,\epsilon 0}$), and structural similarity index ($SSIM$) using the *challengeR* concept [46]. The area of each blob at position (algorithm $i$, rank $j$) is proportional to the relative frequency algorithm $i$ achieved rank $j$, where individual *tasks* (for which rankings are computed) correspond to the solving of the optical inverse problem for different wavelengths. Lower ranks represent better performances. The median rank for each model is indicated by a black cross. The black lines indicate 95% confidence intervals ranging from the 2.5th to the 97.5th percentile.

### 3.1. Forearm data set

When tested on the same annotation-based forearm data set, the data-driven methods clearly outperform the literature-based method that is exclusively based on prior knowledge (cf. Fig. 4). Leveraging the (larger) GAN-based data set also yields a substantial improvement compared to using only the annotations (without further augmentation). These findings hold true irrespective of the specific metric that is applied and in different target structures (cf. Section C.1 in the supplemental material), such as the tissue classes artery, skin, and vein across different wavelengths.

The mean absolute and relative errors of the five competing algorithms (cf. Fig. C.2 in the supplemental material) for both the (different) test sets reflecting the distribution of the respective training set and the joint (most realistic) test set show that the performance is slightly wavelength-dependent. Furthermore, the performance on held-out test data (in distribution) does not generalize to the (more realistic) joint test data for the literature-based method. Additional quantitative results of our comparative performance assessment can be found in Section C.1 in the supplemental material.

Qualitative results for three of the models on the same test data for reasons of visual clarity are shown in Fig. 5. The presented image was chosen according to the median of the mean absolute errors averaged across the whole images at 700 nm ($\overline{AE}_{x,c=700\ nm}$) for the model trained on the literature-based data set. The estimated absorption coefficients differ visually the most from the ground truth coefficients for the models trained on literature-based data. The estimations of the models trained on annotation-based and GAN-based data more closely resemble the ground truth coefficients. However, the absolute error images show that all models perform worse within target structures, such as skin and vessels compared to the background structure.

### 3.2. Calf and neck data set

The ranking stability of the calf model and neck model (cf. Fig. 6) demonstrates that the data-driven models outperform the purely annotation-based models. This finding is again irrespective of the applied metric and the analyzed tissue class. Additional results of our comparative performance assessment can be found in Section C.2 in the supplemental material.

## 4. Discussion

In this work, we proposed a novel framework for photoacoustic image analysis relying on two pillars: (1) the explicit disentanglement of different factors contributing to image formation and (2) the data-driven approach to image synthesis ("learning to simulate"). With the present contribution, we are, to our knowledge, the first to apply an adversarial approach to the simulation of realistic tissue geometries in the specific context of PAT. Our initial feasibility study suggests that the proposed data-driven approach is better suited for PAT tissue synthesis compared to the traditional model-based approach that leverages all the prior knowledge available to optimize the approach.

Although the annotated tissue geometries can be expected to resemble the test data geometries the most due to the same underlying data distribution, the performance of the corresponding model was worse compared to the data-driven methods, independent of the investigated body site. The most plausible explanation for this phenomenon is the small data set size, which motivated our work on synthetic data generation. The fact that the GAN-based forearm methods performed so well compared to the literature-based method indicates that we were able to generate realistic anatomical parameter images in a data-driven manner. While the combination of the three forearm data sets yielded the best performance, additional experiments suggest that this effect can merely be attributed to data set size. In fact, a model trained on a GAN-anno-based data set of the same size as the combined data set was superior to all other models (cf. Fig. C.4 and C.5 in the supplemental material).

It should further be mentioned that the GAN-based approach can essentially be interpreted as a form of data augmentation. Data augmentation in qPAT is in fact non-trivial, as the standard mechanisms (e.g., image translation, rotation, scaling) are not applicable because they would alter the reference annotations of the optical parameters. Here, we overcame this problem by disentangling the geometry from the optical and acoustic properties. In principle, hand-crafted augmentations of the anatomical parameter images can be a valid alternative to our GAN-based approach. However, manual augmentation would require explicit prior anatomical knowledge and thus it would be transferable and generalizable to different applications only with substantial effort. In contrast to hand-crafted augmentations of anatomies, our data-driven augmentation automatically learns the underlying distribution and results in diverse masks that showed a boost in accuracy compared to the annotation-based method relying on the small manually annotated data set.

The quantitative results of the U-Net-based quantification model tested on held-out test data of the different *in silico* data sets (cf. left plots in Fig. C.2 in the supplemental material) confirm the general suitability of this architecture in estimating the absorption coefficient $\mu_a$ from the initial pressure $p_0$ [44,47,48]. The focus of this work was to demonstrate the benefit of our GAN-based concept rather than optimizing the performance of the specific downstream task explored here. Several measures could be taken to further increase segmentation accuracy. It is well known, for example, that ensembling leads to substantial general performance boosts [49]. Similarly, data augmentation can be key to optimizing performance. One data augmentation option would be to generate multiple synthetic PAT images (corresponding to different sampled optical parameters) from the same anatomical parameter image. Additionally, using the off-center slides of the 3D simulated PAT volumes for training of the downstream task could result in a further increase in performance. Furthermore, we only performed the downstream task in 2D because the real data, and thus the manual annotations, were in 2D. Assuming access to real 3D data, we could perform the downstream task in 3D and would expect an improved performance [50].

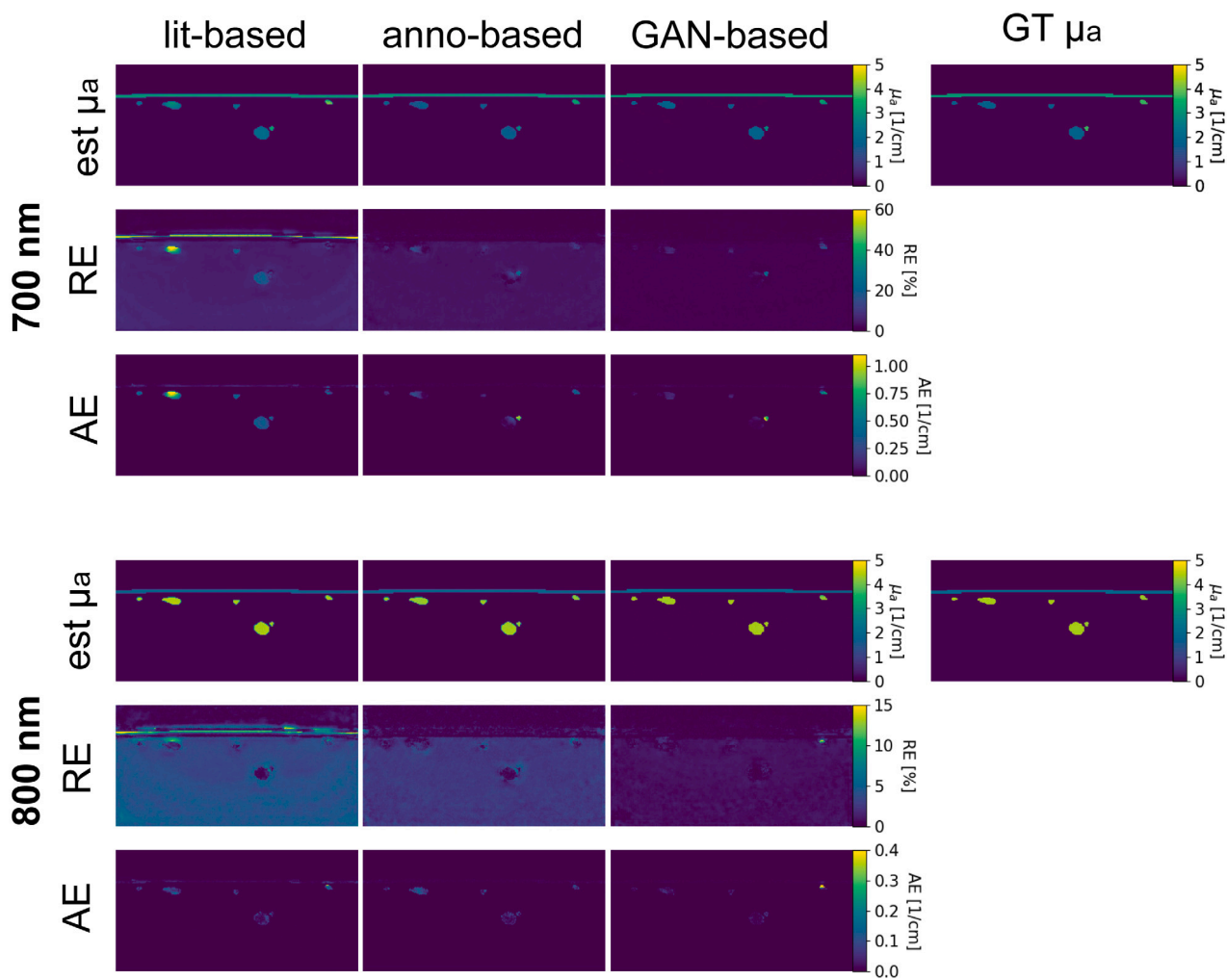

**Fig. 5.** Qualitative results on a representative annotation-based forearm test case. The estimated absorption coefficient (est $\mu_a$), the relative error ($RE$), the absolute error ($AE$), and the corresponding ground truth (GT $\mu_a$) are shown for the models trained on annotation-based (anno), literature-based (lit), and Generative Adversarial Network (GAN)-based forearm data at (*top*) 700 nm and (*bottom*) 800 nm. The image was chosen according to the median mean absolute error ($\overline{AE}$) (more specifically, $\overline{AE}_{x,z=0,\lambda=700\ nm}$ averaged across the whole image) at 700 nm for the model trained on the literature-based data set. The $\mu_a$ estimations of the models including GAN-based data most closely resemble the $\mu_a$ GTs.

All models showed a quantification performance that was slightly dependent on the wavelengths when tested on both in distribution held-out data and annotation-based target test data (cf. Section C in the supplemental material). This behavior might result from inherent wavelength-dependent signal intensity differences and associated different light distributions in the tissue, which are mainly driven by target structures. Moreover, since this dependence was present in target structures, it could be enhanced by the imbalance between the amount of pixels assigned to target classes and other classes. To compensate for this effect, we additionally used the $SSIM$, but the conclusions drawn from the experiments corresponding to the three different metrics were identical.

A point worth considering in this work is that we used a downstream task to investigate the benefit of our learning-to-simulate approach. To get a more complete picture of the strengths and weaknesses of our concept, future work should transfer the method to more applications in a diversity of anatomical regions and classes, including pathologies such as cancer, and explore further downstream tasks. Along these lines, we would like to extend our validation strategy to real data. Our study does not allow us to draw broad conclusions in terms of a clinical setting, as we performed our experiments exclusively in silico. However, validation on real data is by no means trivial to solve because (1) the sim-to-real domain gap is still too large (related to optical and acoustic forward modeling) and (2) realistic experimental setups, in which the underlying properties are precisely known, have to be created. Both issues are subject to ongoing research.

A limitation of our approach could be seen in the fact that we have no guarantee for our anatomical parameter images to accurately reflect tissue geometries. One reason for this is that the annotations were performed on US and PAT data - both modalities that require experience in image interpretation and are highly dependent on the quality of the reconstruction algorithm. However, our general concept is not at all limited to these modalities and could instead be applied to better interpretable images, such as CT or MRI. Furthermore, the optical properties assigned to the parameter images do not reflect the full spectrum of tissue property variations that can occur in practice. Despite these issues, we see the strength of our approach in the fact that we managed to disentangle performance effects that can be attributable to the tissue geometry from those resulting from the optical properties themselves. With this concept, we are the first to show that the performance of DL-based qPAT solutions depends crucially on the method used to generate tissue geometries.

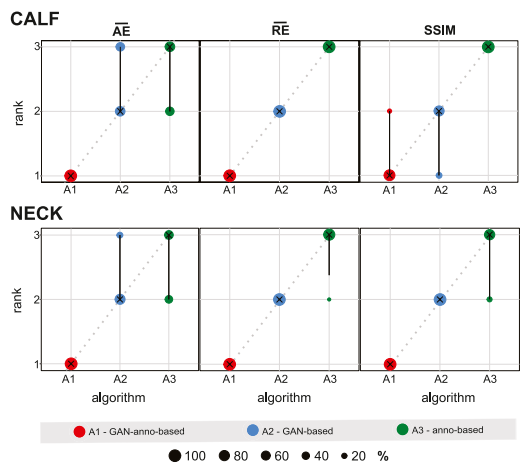

**Fig. 6.** Comparative performance assessment of the *(top)* calf and *(bottom)* neck models corresponding to different training sets (cf. Table 2) and tested on identical annotation-based test data. Uncertainty-aware rankings were computed for the mean absolute error $(\overline{AE}_{c \times 6})$, mean relative error $(\overline{RE}_{c \times 6})$, and structural similarity index ($SSIM$) using the *challengeR* concept [46]. The area of each blob at position (algorithm $i$, rank $j$) is proportional to the relative frequency algorithm $i$ achieved rank $j$, where individual *tasks* (for which rankings are computed) correspond to the solving of the optical inverse problem for different wavelengths. Lower ranks represent better performances. The median rank for each model is indicated by a black cross. The black lines indicate 95% confidence intervals ranging from the 2.5th to the 97.5th percentile.

Overall, we believe that the proposed "learning to simulate" approach has high potential to enhance the realism of synthetic PAT data and could thus become an important concept for generating and augmenting adequate training data for qPAT applications as well as for generating realistic validation data in the field.

**Declaration of competing interest**

The authors declare that they have no known competing financial interests or personal relationships that could have appeared to influence the work reported in this paper.

**Funding**

This project has received funding from the European Research Council (ERC) under the European Union's Horizon 2020 research and innovation programme (grant agreement No. [101002198]) and the Surgical Oncology Program of the National Center for Tumor Diseases (NCT) Heidelberg.

*Ethics approval:*

The healthy human volunteer experiments were approved by the ethics committee of the medical faculty of Heidelberg University under reference number S-451/2020 and the study is registered with the German Clinical Trials Register under reference number DRKS00023205.

**Data availability**

The data that has been used is confidential.

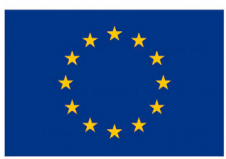

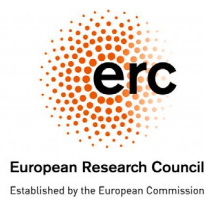

European Research Council
Established by the European Commission

**Appendix A. Supplementary data**

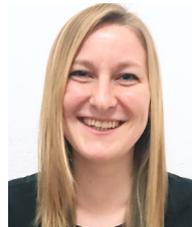

**Melanie Schellenberg** received her M.Sc. degree in Physics from Heidelberg University in 2019. She is currently pursuing an interdisciplinary Ph.D. in computer science at the division of Intelligent Medical Systems (IMSY), German Cancer Research Center (DKFZ) and aiming for quantitative photoacoustic imaging with a learning-to-simulate approach.

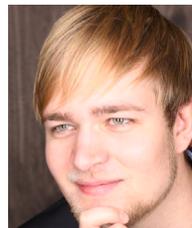

**Janek Groehl** received his M.Sc. degree in medical informatics from Heidelberg University and Heilbronn University of Applied Sciences in 2016. He received his Ph.D. from the medical faculty of Heidelberg University in April 2021. In 2020, he worked as a postdoctoral researcher at the German Cancer Research Center in Heidelberg, Germany and is currently working as a research associate at the Cancer Research UK Cambridge Institute in Cambridge, United Kingdom. He does research in computational biophotonics focusing on data-driven methods for data processing and signal quantification in photoacoustic imaging.

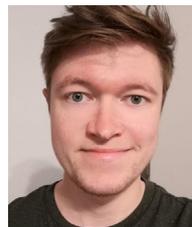

**Kris Kristoffer Dreher** received his M.Sc. degree in Physics from Heidelberg University in 2020. He is currently pursuing a Ph.D. at the division of Intelligent Medical Systems (IMSY), German Cancer Research Center (DKFZ) and does research in deep learning-based domain adaptation methods to tackle the inverse problems of photoacoustic imaging.

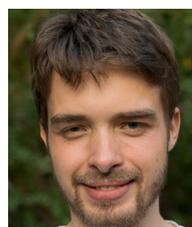

**Jan-Hinrich Nölke** received his M.Sc. degree in Physics from Heidelberg University in 2021. He is currently pursuing an interdisciplinary Ph.D. at the division of Intelligent Medical Systems (IMSY), German Cancer Research Center (DKFZ). His research focuses on deep learning-based uncertainty quantification in photoacoustic imaging.

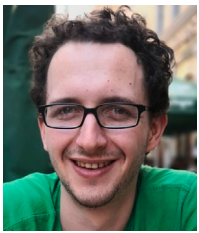

**Niklas Holzwarth** received his M.Sc. degree in Physics from Heidelberg University in 2020. He is currently pursuing an interdisciplinary Ph.D. in computer science at the division of Intelligent Medical Systems (IMSY), German Cancer Research Center (DKFZ) investigating a sensorless 3D photoacoustic approach, referred to as "tattoo tomography".

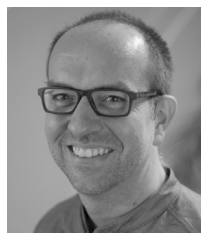

**Alexander Seitel** is a computer scientist currently working as a group lead and deputy head at the division of Intelligent Medical Systems (IMSY) at the German Cancer Research Center (DKFZ) in Heidelberg. He received his Doctorate in Medical Informatics from Heidelberg University and holds a Diploma (M.Sc. equivalent) in Computer Science from the Karlsruhe Institute of Technology. His research focusses on computer-assisted interventions and novel imaging methodologies aiming to improve interventional healthcare. In this area, he conducted various international projects at the DKFZ, during his two-year postdoctoral fellowship at the University of British Columbia, Vancouver, Canada, and at the Massachusetts Institute of Technology (MIT), Cambridge, MA.

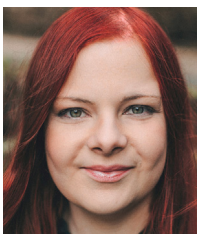

**Minu Dietlinde Tizabi** received her Doctorate of Medicine from Heidelberg University in 2017. She is currently a physician, scientist and writer in the division of Intelligent Medical Systems (IMSY) at the German Cancer Research Center (DKFZ).

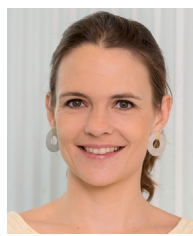

**Lena Maier-Hein** is a full professor at Heidelberg University (Germany) and affiliated professor to LKSK institute of St. Michael's Hospital (Toronto, Canada). At the German Cancer Research Center (DKFZ) she is managing director of the "Data Science and Digital Oncology" cross-topic program and head of the division of Intelligent Medical Systems (IMSY). Her research concentrates on machine learning-based biomedical image analysis with a specific focus on surgical data science, computational biophotonics and validation of machine learning algorithms.